\documentclass[conference]{IEEEtran}
\IEEEoverridecommandlockouts

\usepackage{amsmath}
\usepackage{graphicx}
\usepackage{url}

\usepackage[pdftex]{hyperref}
\usepackage{tikz}
\newcommand\copyrighttext{%
    \footnotesize \textcopyright 2025 IEEE. Personal use of this material is permitted.
    Permission from IEEE must be obtained for all other uses, in any current or future
    media, including reprinting/republishing this material for advertising or promotional
    purposes, creating new collective works, for resale or redistribution to servers or
    lists, or reuse of any copyrighted component of this work in other works.
    DOI: \href{https://doi.org/10.1109/BigData66926.2025.11402160}{https://doi.org/10.1109/BigData66926.2025.11402160}}
\newcommand\copyrightnotice{%
    \begin{tikzpicture}[remember picture,overlay]
        \node[anchor=south,yshift=10pt] at (current page.south) {\fbox{\parbox{\dimexpr\textwidth-\fboxsep-\fboxrule\relax}{\copyrighttext}}};
    \end{tikzpicture}%
}

\newcommand{\figref}[1]{\figurename~\ref{#1}}

\def\BibTeX{{\rm B\kern-.05em{\sc i\kern-.025em b}\kern-.08em
    T\kern-.1667em\lower.7ex\hbox{E}\kern-.125emX}}

\begin{document}

\title{Optimizing Memory Allocation in Distributed Clusters with Predictive Modeling}

\author{\IEEEauthorblockN{
    Jonathan Bader\IEEEauthorrefmark{1},
    Edgar Blumenthal\IEEEauthorrefmark{1},
    Marten Eckardt\IEEEauthorrefmark{1},
    Justus Krebs\IEEEauthorrefmark{1},
    Joel Witzke\IEEEauthorrefmark{1},
    Xemena Wysokinska\IEEEauthorrefmark{1},\\
    Haci Ismail Aslan,
    and Odej Kao}
\IEEEauthorblockA{\textit{TU Berlin, Germany}}
\IEEEauthorblockA{\{jonathan.bader, joel.witzke, aslan, odej.kao\}@tu-berlin.de}
\IEEEauthorblockA{\{e.blumenthal, m.eckardt, justus.krebs, x.wysokinska\}@campus.tu-berlin.de}
}

\IEEEpubid{\makebox[\columnwidth]{*equal contribution, alphabetical order \hfill} \hspace{\columnsep}\makebox[\columnwidth]{ }}

\maketitle
\copyrightnotice

\IEEEpubidadjcol

\begin{abstract}
  In modern distributed systems, efficient resource allocation is a vital aspect to maintain scalability, reduce operational costs, and ensure fast execution even across heterogeneous workloads. Predictive models for resource usage are essential tools for optimizing allocation and preventing system bottlenecks.
  Predictive memory allocation has asymmetric costs as a key challenge: underallocation causes failures while overallocation wastes memory.
  
  We propose a regression method based on a LightGBM and XGBoost ensemble trained to predict high conditional quantiles. To further account for the high cost of underallocations we add a multiplicative safety factor.
  With our method we are able to reduce the number of under-allocated jobs from 4.17\% to 2.89\% and average overallocation from 148\% to 44.51\% on a real-world dataset of build jobs provided by SAP. We further explore the pareto frontier between optimization for underallocation and for overallocation.
\end{abstract}


\section{Introduction}

Distributed clusters are used by many companies and institutions on a daily basis. Because such architectures are the backbone of modern large-scale computing systems, correct resource allocation is a high priority. One such example is memory allocation. This is challenging: underallocation, assigning too little memory, often causes crashes, whereas excessive overallocation wastes resources.

The task of memory allocation has evolved from custom and general-purpose allocators~\cite{gp_allocators}, through dynamic schemes (e.g. binning~\cite{binning}) to machine learning-based optimization~\cite{ml_perf_modeling}.
Other related approaches involve recurrent neural networks~\cite{rnn_prediction}, LSTM based models~\cite{lstm_prediction} and dynamic ensemble learning for scientific workflows~\cite{sizey}.

For this study, we use a dataset from SAP containing detailed traces of build jobs\footnote{\url{https://github.com/SAP/task-execution-data-set}}. Analyzing baseline memory allocation done through manual memory limitations by SAP developers revealed opportunities to reduce both under- and overallocation, motivating the search for and development of a better approach.

We evaluated multiple classification and regression approaches. The best results were achieved using a quantile regression ensemble~\cite{quantile_forests} combining XGBoost~\cite{xgboost} and LightGBM~\cite{lightgbm}. This approach achieved more than a two-thirds reduction in memory waste and reduced underallocation to below 3\%, representing a 50\% improvement over the SAP baseline.

\section{Evaluation}

\subsection{Dataset Features}

The dataset contains Continuous Integration (CI) infrastructure metadata on a cluster. The data includes resource usage information for build jobs. After anonymization, each build has 19 feature columns with the most relevant being:
\begin{itemize}
    \item \emph{time:} timestamp of the build.
    \item \emph{memory{\_}fail{\_}count:} kernel memory allocation failure count during the build; used to identify whether a job was under-allocated.
    \item \emph{buildProfile:} contains informaton about the architecture, compiler, and the optimization level.
    \item \emph{makeType:} build type indicator.
    \item \emph{max{\_}rss:} peak memory usage during the build (bytes).
    \item \emph{memreq:} requested memory for the build container (MB).
\end{itemize}

We derived an additional 21 predictive features from the initial 19 raw telemetry columns through feature engineering (excluding one-hot encodings).
We applied temporal decomposition to capture daily and weekly seasonality in resource usage~\cite{temporal_forecasting}, build profile parsing, workload characterization and lagged statistical history.

After creating our model, the five most important features were \emph{lag{\_}1{\_}grouped} (memory usage of the direct predecessor job), \emph{rolling{\_}p95{\_}rss{\_}g1{\_}w5} (rolling 95th percentile of memory usage for similar jobs), \emph{jobs}, \emph{branch{\_}id{\_}str} and \emph{ts{\_}weekofyear}. This dominance of historical features validates our feature engineering strategy and demonstrates that a job's recent, localized memory history is the most critical predictor for its future memory needs.

\subsection{Hyperparameter Search}

We used Bayesian optimization to tune each model's hyperparameters using the \emph{Optuna} library~\cite{optuna}. Bayesian optimization builds a surrogate model of performance (e.g., a Gaussian process) and selects new hyperparameters via an acquisition policy (e.g., expected improvement). Here, the TPE sampler was used~\cite{optuna_tpe} in accordance to Snoek et al.~\cite{bayesian_optimization}, balancing high predicted performance (exploitation) against uncertainty (exploration) and makes use of past evaluations for guidance, often finding good hyperparameter settings in fewer trials than simple grid or random search.

The search space included generic parameters (e.g. learning rate, number of trees/estimators) and model-specific parameters such as the quantile parameter for the ensembles.
The cost function used for training the models balances the amount of under-allocated jobs (\%) against the total overallocation ratio to ensure predictions are sufficient but not wasteful:
$$
\min_\theta Cost(\theta) = 5 \cdot \text{under\_alloc}(\theta) + \frac{\text{total\_predicted\_rss}(\theta)}{\text{total\_actual\_max\_rss}}
$$
where $\theta$ are the parameters of the respective model.
We penalized underallocation errors since they could lead to job failures, wasting previous computation and necessitating a rerun.
The penalty factor was set to 5 based on empirical tuning and represents our trade-off between prioritizing job success while maintaining overall resource efficiency.

Each trial consisted of 3-fold cross-validation to ensure that the chosen hyperparameters generalize across folds and do not overfit a single train/validation split. We pooled all out-of-fold (OOF) predictions and their corresponding true values from each fold to calculate $Cost(\theta)$ once on the aggregated predictions. This gives equal weight to every test instance, providing a more robust performance estimate than averaging the costs of each fold~\cite{cross_validation}, since the distribution of target values and the size of each fold can vary significantly.

\subsection{Regression Ensemble Setup}

To further reduce underallocation, we predicted the upper $\alpha$-quantile (with $\alpha \in [0.90, 0.99]$) of memory instead of the mean. This approach trains the model to produce predictions that are expected to be higher than the true value in $\alpha$\% of cases, naturally reducing the risk of underallocation~\cite{quantile_regression}.

We evaluated both single quantile regressors and multiple two-model quantile ensembles. Ensembles of models can reduce correlation between learners and often generalize more robustly on tabular data~\cite{ensemble_selection}.
Each of the base learners $a$ and $b$ predicts an $\alpha$-quantile, and the ensemble output is the per-row maximum, scaled by a safety factor $s \in [1.00, 1.15]$:
$$
\hat{y}(x) = \max \left(\hat{y}^{(a)}_\alpha (x), \hat{y}^{(b)}_\alpha (x)\right) \cdot s
$$

The per-row maximum aggregation method was selected for its simplicity and effectiveness in enforcing a conservative allocation strategy. This approach ensures that for any given job, the allocation is guided by the more cautious of the two models, thereby minimizing the risk of underallocation. While we explored other techniques, such as mean and percentile-based aggregation~\cite{quantile_aggregation}, as well as expanding the ensemble with more models, these alternatives failed to deliver better performance and introduced unnecessary complexity. Therefore, maximum aggregation with two quantile loss models provided the optimal balance of caution, accuracy, and simplicity.

In our regression experiments we focused on gradient-boosted tree models: Scikit-learn's GradientBoosting (GB), XGBoost (XGB), LightGBM (LGB), and CatBoost (Cat). We evaluated the following model families:
\begin{itemize}
    \item \textbf{Heterogeneous Ensembles:} Various pairwise combinations of the four base learners (e.g., LGB+XGB, GB+LGB).
    \item \textbf{Homogeneous Ensembles:} Combinations of two identical learners, each with an independent hyperparameter search (e.g., XGB+XGB, Cat+Cat).
    \item \textbf{Single-Model Baselines:} Standalone LightGBM and XGBoost models using their native quantile objectives.
\end{itemize}
Heterogeneous pairs (e.g., LGB+XGB) provide diversity across libraries, which reduces error correlation without leaving the quantile framework. This focus also aligns with benchmarks showing that gradient-boosted models are consistently state-of-the-art for structured tabular data~\cite{tree_based_outerperformance, deep_learning_unneeded}.

\subsection{Exploring the Pareto Frontier and Results}

\begin{figure}
  \centering
  \includegraphics[width=\linewidth]{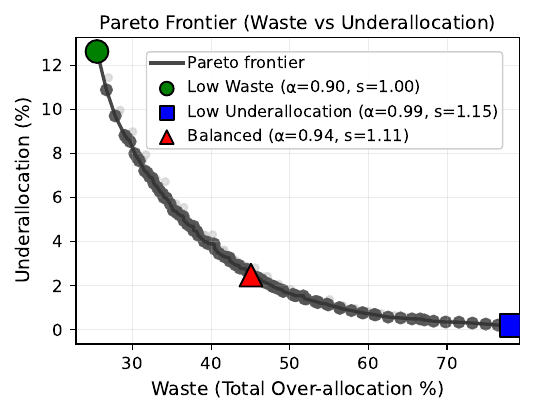}
  \caption{Pareto frontier for our LightGBM+XGBoost ensemble model with points corresponding to different ($\alpha$, $s$) pairs.}
  \label{fig:pareto_frontier}
\end{figure}

By adjusting the model's quantile level ($\alpha$) and safety multiplier ($s$), we can create different versions of the model to prioritize either safety or efficiency.
This creates a pareto frontier between optimizing for under- vs. for overallocation shown in \figref{fig:pareto_frontier}. Having different models on that frontier allows for flexibility and can serve different operational needs or business priorities:
\begin{itemize}
    \item \textbf{Balanced:} Our best performing model based on our defined cost function, offering a strong balance between safety and efficiency (2.89\% underallocation, 44.51\% overallocation).
    \item \textbf{Low Waste:} An aggressive model that minimizes overallocation (25.5\%) at the cost of a higher underallocation rate (12.6\%).
    \item \textbf{Low Under-allocation:} A very conservative model that nearly eliminates underallocation (0.20\%) at the cost of higher resource waste (78.2\%). That is still substantially lower than the baselines 148\% on the same hold-out set.
\end{itemize}

\begin{figure}
  \centering
  \includegraphics[width=\linewidth]{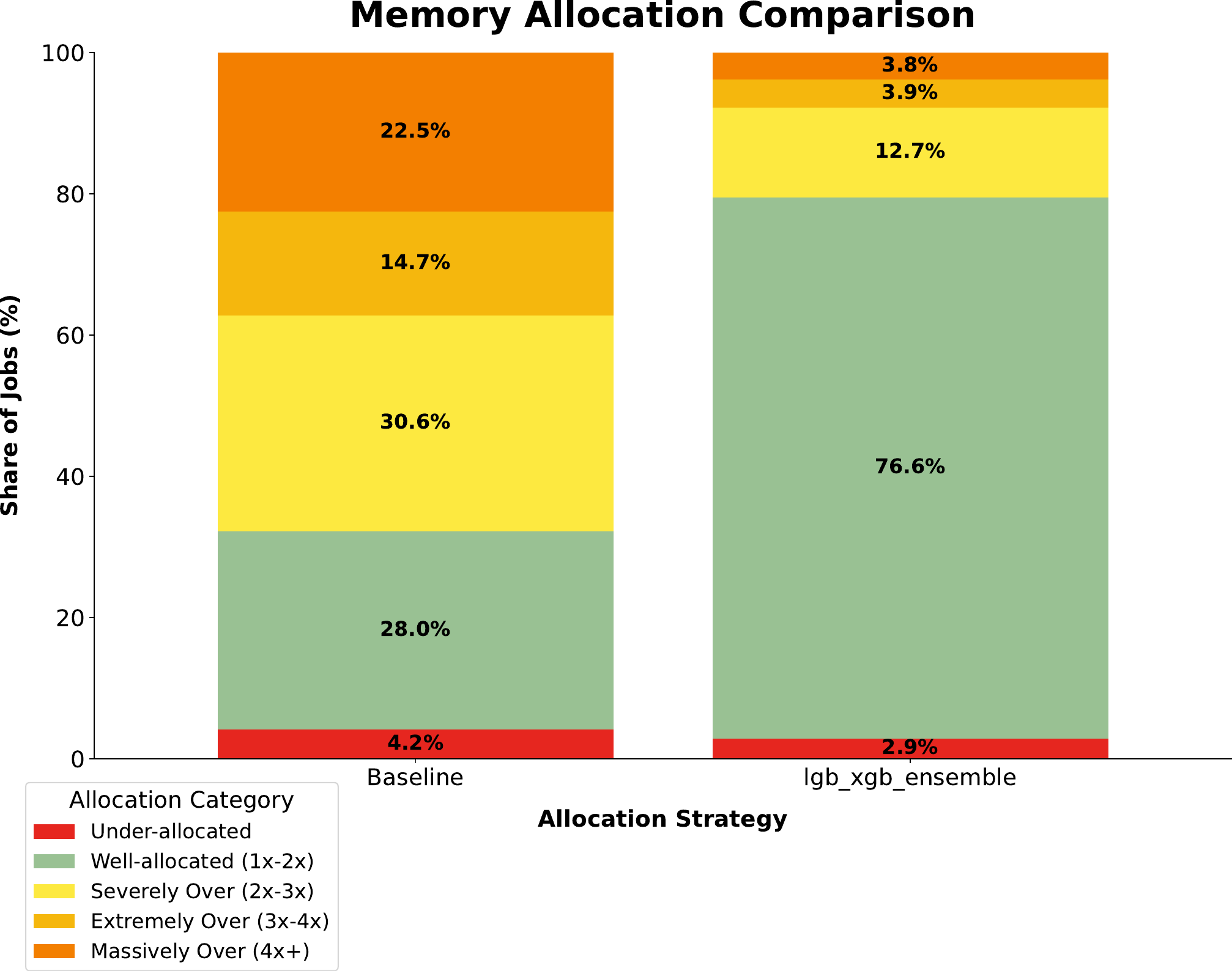}
  \caption{Job distribution by allocation quality for our method vs. the baseline.}
  \label{fig:topmodels}
\end{figure}

\figref{fig:topmodels} shows the performance in different levels of overallocation of our balanced approach compared to the baseline.
With inference taking about $2\cdot10^{-5}$ seconds it is also suitable for predictions in realtime scenarios.

\section{Conclusion}

In this paper we addressed the challenge of memory allocation dealing with asymmetric cost as under-provisioning is significantly more costly than over-provisioning. Our proposed quantile regression ensemble reduced underallocation from the baseline's 4.17\% to 2.89\% and, critically, decreased total overallocation (resource waste) by over 70\% from 148\% to 44\%.
A key finding from our experiments is that a conservative architectural design was more influential than the specific choice of algorithm. The consistent high performance across different regression ensemble combinations indicates that the core strategy of using a high-quantile objective and max aggregation was the primary driver of success.

Future work could extend this research, e.g., by exploring a hybrid policy that uses an additional conservative model for build jobs predicted to require large amounts of memory, where underallocation implications are the highest.
Another approach could use reinforcement learning to learn an optimal allocation online, dynamically adapting to changing workloads and current cluster utilizations in real-time.

Our code, including an application for realtime predictions and trained models, can be found on GitHub\footnote{\url{https://github.com/Zmart64/SAPResourceOptimizer}}.

\section*{Acknowledgments}
\thanks{This paper was partially supported by the Swarmchestrate project of the European Union's Horizon 2023 Research and Innovation programme under grant agreement no. 101135012.}

\thanks{Further funding received by the Deutsche Forschungsgemeinschaft (DFG, German Research Foundation) as FONDA (Project 414984028, SFB 1404).}

\thanks{We also thank SAP for their data and friendly cooperation.}

\bibliographystyle{IEEEtran}
\bibliography{literature}

\end{document}